\documentclass[prb,aps,twocolumn,showpacs]{revtex4}
\usepackage{graphicx,dcolumn,amsmath,color,longtable}

\begin{document}

\title{Giant magnetic anisotropy of the bulk antiferromagnets IrMn and IrMn$_3$} 

\author{L.\ Szunyogh$^1$}\email{szunyogh@phy.bme.hu}
\author{B.\ Lazarovits$^{1,2}$}
\author{L.\ Udvardi$^1$}
\author{J.\ Jackson$^3$}
\author{U.\ Nowak$^4$}

\affiliation{$^1$ Department of Theoretical Physics, Budapest University of Technology and Economics, Budafoki \'ut 8. H1111 Budapest}
\affiliation{$^{2}$Research Institute for Solid State Physics and Optics, Hungarian Academy of Sciences, H-1525 Budapest, PO Box 49, Hungary}
\affiliation{$^3$ Department of Physics, University of York, York YO10 5DD, United Kingdom}
\affiliation{$^4$ Fachbereich Physik, Universit\"at Konstanz, 78457 Konstanz, Germany}

\date{\today}

\begin{abstract}
Theoretical predictions of the magnetic anisotropy of antiferromagnetic materials are demanding 
due to a lack of experimental techniques which are capable of a direct measurement of this quantity.
At the same time it is highly significant due to the use of antiferromagnetic components 
in magneto-resistive sensor devices where the stability of the antiferromagnet is of 
upmost relevance.  We perform an ab-initio study of the ordered phases of IrMn and IrMn$_3$, 
the most widely used industrial antiferromagnets.  Calculating the form and the 
strength of the magnetic anisotropy allows the construction of an effective spin model, 
which is tested against experimental measurements regarding the magnetic ground state 
and the N\'eel temperature.  Our most important result is 
the extremely strong second order anisotropy 
for IrMn$_3$ appearing in its frustrated triangular magnetic ground state, 
a surprising fact since the ordered $L$1$_2$ phase has a cubic symmetry.  
We explain this large anisotropy by the fact that cubic symmetry is locally broken 
for each of the three Mn sub-lattices.
\end{abstract}


\maketitle

While the magnetic anisotropy (MA) of ferromagnets is a well investigated quantity, 
both experimentally as well as theoretically, it is much less understood in case of 
antiferromagnets.  This lack of knowledge is on the one hand due to a lack of experimental 
techniques which are capable of a direct measurement of this quantity.  On the other hand, 
theoretical first principles calculations of magnetic anisotropy effects are quite challenging 
as they require the use of fully relativistic spin density functional theory. 

Interest in the MA of antiferromagnets comes from the fact that these compounds
are important components of GMR sensors used, e.g., in hard disc read heads.  
Antiferromagnetic materials are employed in these devices to form antiferromagnet/ferromagnet 
bilayers exhibiting exchange bias\cite{noguesJMMM99}, a shift of the hysteresis loop of the 
ferromagnet, providing a pinned layer which fixes the magnetization of the reference layer of 
a GMR sensor.  The stability of the antiferromagnet is most crucial for the stability of 
exchange bias and hence the functioning of the device \cite{miltenyiPRL00,nowakJMMM02}.  Industrially the antiferromagnet IrMn is widely used because of the large exchange bias and thermal stability that can be obtained with this material.

From experimental investigations of the exchange bias effect it is concluded that IrMn must 
have a rather large MA.  Recent estimates of the MA of IrMn concerned the mean blocking temperature $T_B$, the temperature at which the exchange bias shift changes sign upon thermal activation.  From T$_B$ the intrinsic MA can be inferred if the particle size distribution is known; such a procedure has recently been reported and the room temperature MA energy of IrMn was estimated at 
5.5$\times10^6$erg/cc~\cite{vallejoAPL07} and even 2.8$\times10^7$erg/cc~\cite{aleyIEEE09} depending on the seed layer and, consequently, the texture of the IrMn. 

In this letter, we address several features of the MA of IrMn alloys starting from first principles.  
In terms of simple symmetry considerations we predict the form of the MA energy
that we fully confirm using ab-initio calculations providing also the strength of the MA,
i.e., the relevant MA constants. To our best knowledge, for frustrated antiferromagnets,
such as IrMn$_3$, this is the first theoretical prediction of the MA in the literature.
Our most remarkable observation is the surprisingly strong, second order MA of IrMn$_3$ resulting 
from the fact that the cubic symmetry is locally broken for each of the three Mn sub-lattices. 
We are also able to attribute contributions of the MAE related to on-site and 
two-site exchange anisotropy terms, a very important issue for finite 
temperature magnetism\cite{mryasovEL05,stauntonPRB06}. 
Such a separation is inevitably important for the purpose of subsequent 
simulations to study exchange-bias systems based on these compounds, 
for example in determining the scaling behavior of the MA energy.

Self-consistent calculations are performed in terms of the fully relativistic screened 
Korringa-Kohn-Rostoker (SKKR) method \cite{szunyoghPRB95}.  Within this method, spin-polarization and relativistic effects, in particular, spin-orbit coupling are treated on equal theoretical footing by solving the Kohn-Sham-Dirac equation. The local spin--density approximation as parametrized by Vosko \emph{et al. }\cite{voskoCJP80} was applied; the effective potentials and fields were treated within the atomic sphere approximation with an angular momentum cut--off of $\ell_{max}=2$.  For the $L$1$_0$ IrMn alloy we used the lattice constants $a=3.855$\AA~and $c=3.644$\AA\cite{umetsuPRB04}, while for the $L$1$_2$ IrMn$_3$ alloy an fcc lattice with $a=3.785$\AA~was considered \cite{tomenoJAP99,sakumaPRB03}.  
For the self-consistent calculations we fixed the orientations of the magnetic moments 
on the Mn atoms according to the magnetic ground-state configurations reported previously 
in the literature, namely, a checkerboard collinear AF structure for $L$1$_0$ 
IrMn \cite{selteACS68,umetsuPRB04} and a triangular ($T1$) state within the fcc(111) planes 
for $L$1$_2$ IrMn$_3$ \cite{tomenoJAP99,sakumaPRB03}.  
We obtained vanishing spin-polarization at the Ir sites, whereas spin magnetic moments 
of 2.63~$\mu_B$ and 2.66~$\mu_B$ at the Mn sites for IrMn and IrMn$_3$, 
respectively.
These values are in satisfactory agreement with earlier first principles 
calculations~\cite{umetsuPRB04,sakumaPRB03}.

We start our study of the MA by symmetry considerations based on the following
effective spin-Hamiltonian (energy per unit cell), 
\begin{eqnarray}
H  &=&
-\frac{1}{2}\sum_{a,b=1}^{n} J_{ab}\vec{S}_{a}\vec{S}_{b} 
-\frac{1}{2}\sum_{a,b=1}^{n}\vec{S}_{a} \mbox{\boldmath $D$}_{ab} \vec{S}_{b} \nonumber \\
&& -\sum_{a=1}^{n}\vec{S}_{a} \mbox{\boldmath $K$}_{a}\vec{S}_{a}\;,
\label{eq:heisanis}
\end{eqnarray}
where $\vec{S}_a$ is the spin-vector of the Mn sub-lattice labeled by $a$; $n=2$ 
for $L$1$_0$ IrMn and $n=3$ for $L$1$_2$ IrMn$_3$.  $\mbox{\boldmath $D$}_{ab}$ are 
(traceless) symmetric matrices representing anisotropic two-site (exchange) coupling 
and $\mbox{\boldmath $K$}_{a}$ are on-site anisotropy matrices.\cite{udvardiPRB03}  
Note that all the parameters in Eq.~(\ref{eq:heisanis}) are defined as sums over sites 
in the sub-lattices, e.g., $J_{ab} = \sum_{j \in b} J_{ij}$ for $i \in a$ ($j = i$ excluded),
$J_{ij}$ being the isotropic intersite interactions.  
In case of $L$1$_0$ IrMn, tetragonal symmetry implies,
\begin{equation}
\mbox{\boldmath $D$}_{ab} = D_{ab} 
\left(
\begin{array}
[c]{ccc}%
-\frac{1}{2} & 0 & 0\\
0 & -\frac{1}{2} & 0\\
0 & 0 & 1
\end{array}
\right) \; , \quad 
\mbox{\boldmath $K$}_{a} = K
\left(
\begin{array}
[c]{ccc}%
0 & 0 & 0\\
0 & 0 & 0\\
0 & 0 & 1
\end{array}
\right) \; , 
\label{eq:matrices-z}
\end{equation}
with $D_{11}=D_{22}=D$ and $D_{12}=D^\prime$.  Rotating an antiferromagnetic configuration 
around the (100) axis, $\vec{S}_{1}=(0,\sin\varphi,\cos\varphi)$ and $\vec{S}_{2}=-\vec{S}_{1}$, 
a simple orientation ($\varphi$-) dependence of the energy can be derived, 
$E(\varphi) = E(0) + K_{\mathrm{eff}} \sin^2\varphi$, introducing an effective uniaxial MA constant per unit cell, $K_{\mathrm{eff}} = 2K+\frac{3}{2} (D^\prime -D )$.  

In order to calculate $E(\varphi)$ from first principles we adopted 
the so-called {\em magnetic force theorem}~\cite{jansenPRB99} in which the previously 
determined self-consistent effective potentials and fields are kept fixed and 
the change of total energy of the system with respect to $\varphi$ is approached by that of 
the single-particle (band-) energy.  The values for $E(\varphi)$ from these calculations 
could be very well fitted with $K_{\mathrm{eff}}=-6.81$meV, 
in very good agreement with the 
theoretical value reported by Umetsu {\em et al.} \cite{umetsuAPL06} and also with 
the easy-plane anisotropy observed experimentally \cite{selteACS68}.  
Furthermore, the on-site MA constant $K$ in Eq. (\ref{eq:matrices-z}) can be expressed as, 
\begin{equation}
K = - \frac{1}{2} \left( 
\left.  \frac{\partial^2 E}{\partial \theta_i^2} \right|_{(100)} -
\left.  \frac{\partial^2 E}{\partial \phi_i^2} \right|_{(100)} 
\right) \: ,
\label{eq:onsite}
\end{equation}
where $i$ labels any Mn site (see Ref.~\onlinecite{udvardiPRB03} for details).  Importantly, an overall collinear magnetic arrangement along the $(100)$ axis has to be considered in these calculations.  By using the above formula we obtained $K=-2.94$~meV implying that in this system 
the MA energy is dominated by the on-site anisotropy, 
i.e.~the third term in Eq.~(\ref{eq:heisanis}).

\begin{figure}[ht]
\begin{center}
\includegraphics[width=6cm]{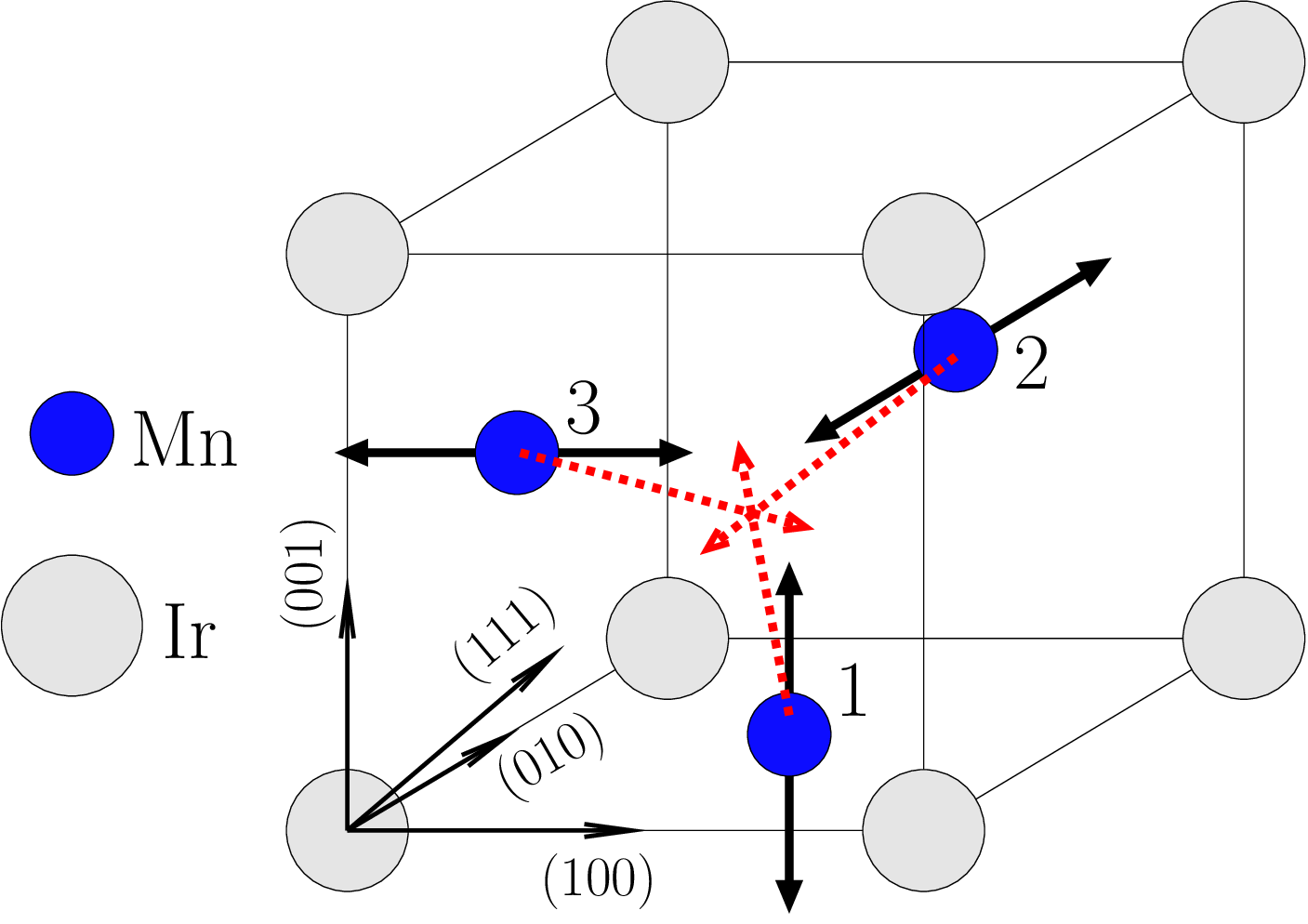}
\end{center}
\caption{(color online) Sketch of the IrMn$_3$ unit cell. 
Dark spheres represent three Mn atoms corresponding to
 the antiferromagnetic sub-lattices. The solid arrows indicate the local easy axes and the dotted arrows indicate the spin direction in the T1 ground state.}
\label{d:systemdiag}
\end{figure}

In the case of $L$1$_2$ IrMn$_3$ each of the three Mn atoms in a unit cell exhibits 
a second order MA due to local tetragonal symmetry.  However, as indicated in Fig.~\ref{d:systemdiag} 
with different symmetry axes that have to be accounted for in Eq.~(\ref{eq:heisanis}) 
by suitable transformations of the matrices in Eq.~(\ref{eq:matrices-z}).  
$C_3$ rotational symmetry around the (111) axes implies $D_{11}=D_{22}=D_{33}=D$ 
and $D_{12}=D_{23}=D_{31}=D^\prime$.  Clearly, for a ferromagnetic state of the system 
such a Hamiltonian would yield a vanishing MA energy. 

\begin{figure}[ht]
\begin{center}
\includegraphics[width=8cm]{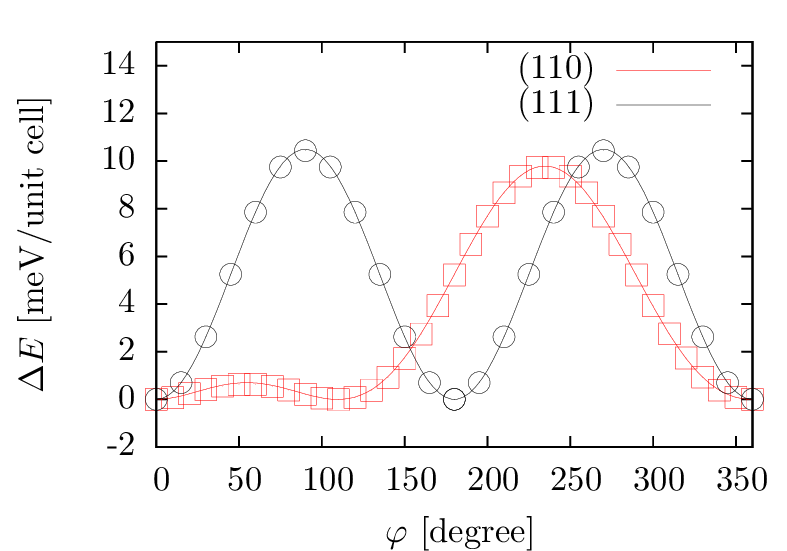}
\end{center}
\vskip -12pt
\caption{(color online) 
Calculated change of energy of the $L$1$_2$ IrMn$_3$ system when rotating
the triangular $T1$ spin structure around the (111) axis (circles) 
and the (110) axis (squares).
The solid lines display appropriate fits to $K_{\rm eff} \sin^2(\varphi)$ 
and the function in Eq.~(\ref{eq:EL12B}), respectively.}
\label{fig:mn3ir-mae}
\end{figure}

This second order MA becomes, however, evident if all the spins forming the $T1$
ground-state are rotated around the (111) axis.  Straightforward calculations show that 
$E(\varphi)$ follows  again a $\sin^2\varphi$--dependence with an effective MA constant, 
$K_{\mathrm{eff}} = 2K+\frac{3}{2} (D+D^\prime)$.  Our first principles calculations reproduced 
well the proposed functional form of $E(\varphi)$ 
with a value of $K_{\mathrm{eff}}=10.42$ meV, see Fig.~\ref{fig:mn3ir-mae}.
Thus we conclude that 
the MA constant for $L$1$_2$ is almost twice as large in magnitude than for $L$1$_0$ IrMn. 

We confirm the validity of the spin-Hamiltonian Eq.~(\ref{eq:heisanis}) for $L$1$_2$ IrMn$_3$ 
by applying two additional rotations of the spin-system. First, we repeat the rotation 
around the (111) axis by simultaneously interchanging the orientations  of the spins
at the Mn sites 2 and 3.  It should be mentioned that this triangular spin-structure (say, $T2$) 
corresponds to a chirality vector,
\begin{equation}
\vec{\kappa} = \frac{2}{ 3 \sqrt{3}} \left( \vec{S}_1 \times \vec{S}_2  
+ \vec{S}_2 \times \vec{S}_3 + \vec{S}_3 \times \vec{S}_1 \right) \; ,
\end{equation}
that is just the opposite of the chirality vector related to state $T1$.  Note also that 
$\vec{\kappa}$ is normal to the plane of the moments and the normal component 
of this vector (chirality index) $\kappa$ for state $T1$ is $\kappa=1$, while for state 
$T2$ \ $\kappa=-1$.  Whilst by considering only the first (isotropic) 
term in Eq.~(\ref{eq:heisanis}) 
the energy of the these two states is identical, the anisotropy terms lift this degeneracy.  
Interestingly, rotating the spins in state $T2$ around the (111) axis does not 
induce changes in the energy of the system. This is confirmed by our 
calculations up to an absolute error of 2~$\mu$eV. 
Furthermore, the energy of state $T2$ should be higher by 
$-K_{\mathrm{eff}}/2$ than the energy minimum of state $T1$ ($\varphi=0$).  
From our calculations we found this difference to be $5.22$~meV, 
fitting nearly perfectly to the previously determined MA constant.

Our last test to Eq.~(\ref{eq:heisanis}) referred to rotating the spins in state $T1$ 
around the (110) axis. As compared to all the previous cases, \
this rotation implies a quite complicated form of $E(\varphi)$, 
\begin{eqnarray}
E(\varphi) &=& E(0) + \frac{K_{\mathrm{eff}}}{8} \Big( 2 + \sin^2\varphi 
-2 \cos\varphi \nonumber \\
&& -2\sqrt{2} \sin\varphi (1-\cos\varphi) \Big) \; .
\label{eq:EL12B}
\end{eqnarray}
In Fig.~\ref{fig:mn3ir-mae} we also plotted the results of this
calculation together with the fit function as above.  
Reassuringly, this function describes $E(\varphi)$ well for the whole range 
of $\varphi$ with the MA constant as obtained before 
($K_{\mathrm{eff}}$=10.42~meV). 
Note that for the rotation around the (110) axis at $\varphi=109.47^\circ$ 
the energy of the ground-state is regained.  
This, however, is not surprising since by this rotation we obtain a $T1$ state 
lying in a plane normal to the (1$\overline{1}\overline{1}$) direction.

In order to calculate the on-site anisotropy parameter, $K$, 
we again applied Eq.~(\ref{eq:onsite}) by using a ferromagnetic 
reference state oriented along the (100) direction.
Although we used the effective potentials 
calculated from the $T1$ ground-state, 
because of the large difference in the spin-configuration 
of the reference state and the true ground-state, 
we expect just a rough estimate on $K$.  
The obtained value, $K \simeq 1.06$ meV, indicates at best that, unlike the $L$1$_0$ IrMn alloy, in this system the MAE is mainly governed by two-site anisotropy, i.e., the second 
term in Eq.~(\ref{eq:heisanis}).

In the second part of this Letter we present results of finite temperature simulations on the
magnetism of IrMn compounds. With this purpose we construct a simplified effective spin model based on our first principles calculations,  
\begin{equation}
H = - \frac{1}{2} \sum_{i \ne j} J_{ij} \vec{S}_i \vec{S}_j  - k \sum_i ( \vec{S}_i \cdot \vec{n}_i)^2.
\label{eq:hamsim}
\end{equation}
where $J_{ij}$ are isotropic Heisenberg exchange parameters and the effective on-site 
anisotropy parameters $k = K_{\mathrm{eff}}/2$, 
 merging thus the effect of two-site anisotropy terms. 
Here, $\vec{ n}_i$ are unit vectors along the local uniaxial symmetry axes:
for $L$1$_0$ IrMn $\vec{ n}_i$ is perpendicular to the Ir (or Mn) planes, 
for $L$1$_2$ IrMn$_3$ $\vec{n}_i$ is different for each of the three Mn sub-lattices, 
see Fig.~\ref{d:systemdiag}.
We calculated the parameters, $J_{ij}$,
by using the widely adopted torque method\cite{lichtensteinJMMM87} 
as extended to relativistic calculations.\cite{udvardiPRB03} 

\begin{figure}[ht!]
\begin{center}
\includegraphics[width=7.5cm]{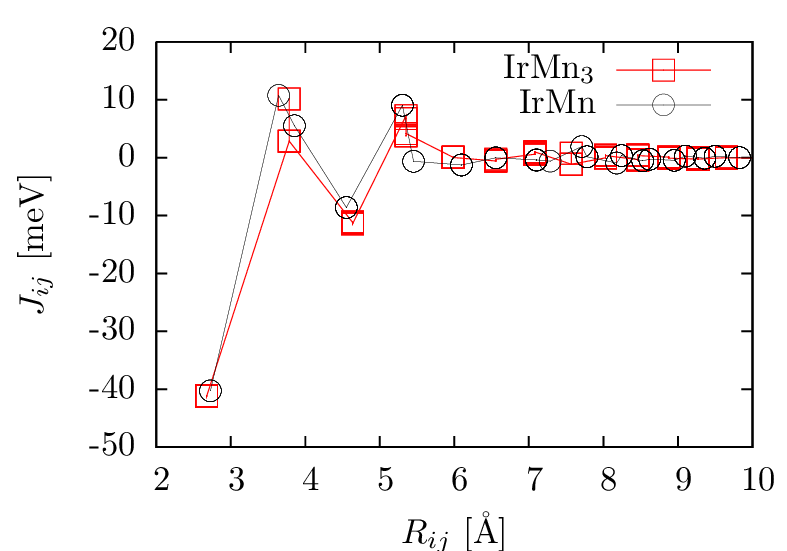}
\end{center}
\vskip -20pt
\caption{(color online) Isotropic exchange interactions, $J_{ij}$, between the Mn atoms 
in IrMn alloys calculated from the corresponding ground state 
magnetic configurations by using the torque method.\cite{udvardiPRB03}}
\label{b:j0fig}
\end{figure}

For both alloys, the calculated exchange interactions are shown in Fig.~\ref{b:j0fig} 
as a function of the distance between the Mn atoms. The two sets of interactions 
show obvious similarities: large antiferromagnetic (negative) nearest neighbor interactions, 
sizable oscillating interactions up to about $R_{ij}=6$~\AA, while 
a strong damping for larger distances.
Note that double (multiple) values for some $R_{ij}$'s  
appear due to the different symmetry (neighborhood) of pairs with the given separation.  
In case of $L$1$_0$ IrMn these 'degeneracies' are mostly resolved via tetragonal distortion 
of the lattice.  In good comparison with other theoretical works~\cite{umetsuPRB04,sakumaPRB03} from the calculated $J_{ij}$'s the mean-field estimates for the N\'eel temperatures, $T_N=1398$K and $1222$K, can be obtained, respectively. 

The model Eq. (\ref{eq:hamsim}) is simulated by solving the Landau-Lifshitz-Gilbert (LLG)
equation with Langevin dynamics, calculating thermal equilibrium properties 
in the long time (and high damping) limit. The methods we use are described in detail 
in Ref. \cite{nowakBOOK07}.  The main quantity of interest is the sub-lattice 
staggered magnetization, $M_s$, defined as
\begin{equation}
M_{s}=\frac{1}{n} \sum_{a=1}^{n} \left\langle \sqrt{ M_{ax}^2 + M_{ay}^2 + M_{az}^2} \right\rangle ,
\end{equation}
where $\vec{M_a} = \sum_{i \in a} \vec{S_i}$ is proportional to the magnetization 
of sub-lattice $a$, $n$ is the number of antiferromagnetic sub-lattices and 
$\langle \; \rangle$ denotes a thermal average.

\begin{figure}[ht!]
\begin{center}
\includegraphics[angle=90, width=8.0cm]{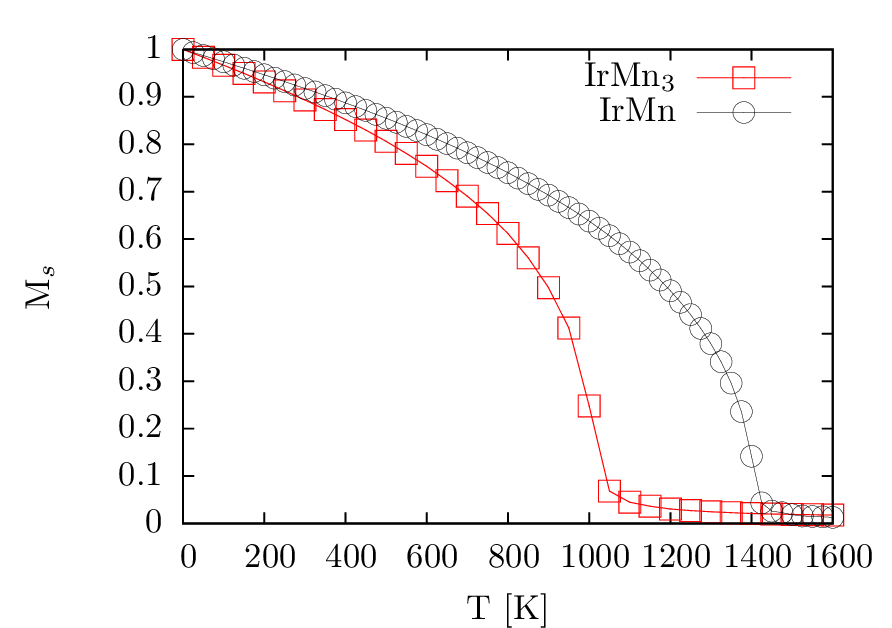}
\end{center}
\vskip -12pt
\caption{(color online) Staggered magnetizations, $M_s$, as a function of temperature
obtained using Langevin dynamics over 20~ps with system sizes of 24000 sites 
($L$1$_2$) and 70000 sites ($L$1$_0$) and using
periodic boundary conditions. }
\label{a:neeltemp}
\end{figure}

Fig.~\ref{a:neeltemp} shows the order parameter, $M_s$, versus temperature $T$. 
Despite finite size effects, $T_N$ can be estimated as 1360K for $L$1$_0$ IrMn and 1005K 
for $L$1$_2$ IrMn$_3$. Note that though the exchange parameters in both cases have similar values, 
the critical temperature in the $L$1$_2$ phase is significantly lower.  Obviously, the frustration 
of the spin-ordering in the $L$1$_2$ phase leads to a reduced T$_N$ as compared 
to the $L$1$_0$ phase.  The simulated critical temperatures clearly 
improve upon the mean field estimates as compared with experimentally observed N\'eel temperatures, 
1145K and 960K\cite{tomenoJAP99}, respectively.  

A further analysis of the sub-lattice magnetization vectors reveals the magnetic  
ground state configurations. In the case of $L$1$_0$ IrMn the Mn spins align along 
the (110) direction appropriate with the easy plane anisotropy for this material.  
For the $L$1$_2$ system, magnetic anisotropy included according Eq.~(\ref{eq:hamsim}) 
reveals that the $T1$ ground state structure is fixed to lie in one of the (111) planes, with each of the Mn spins directed along the corresponding (2$\overline{11}$) directions.  These spin-orientations have previously been established by neutron scattering\cite{tomenoJAP99}; 
our results for the N\'eel temperature and the magnetic ground-state structures 
are in excellent agreement with experiments, underpinning the validity of our spin model
derived from first principles.

In summary, we performed an ab-initio study for the ordered phases of IrMn and IrMn$_3$,   
the most important industrial antiferromagnets. The calculated Heisenberg 
exchange integrals and
magnetic anisotropy constants are used to construct an effective spin model which is 
simulated using the stochastic Landau-Lifshitz-Gilbert equation. 
A good agreement of the calculated N\'eel temperatures and magnetic ground-states 
with experimental results confirmed the validity of our approach.  
Our most spectacular finding is a giant
second order magnetic anisotropy for IrMn$_3$, leading to energy barriers of 
the order of $K_{\mathrm{eff}} \simeq 3\times 10^8$~erg/cc 
for rotation of the T1 ground state spin-structure around the (111) axis.
This uniaxial magnetic anisotropy is understood due to the fact 
that the cubic symmetry 
is locally broken for each of the three sub-lattices of the antiferromagnetic $T1$ ground-state.   

The extremely high anisotropy for the $L$1$_2$ phase has perhaps not been measured 
experimentally because of the disordered nature of this material in thin film devices, 
where deposition by sputtering causes significant loss of long range crystallographic order.   
Our results, however, suggest that finer control of the crystallography will allow the 
extremely large anisotropy of these materials to be fully exploited, 
allowing, for example, antiferromagnet film thicknesses to be reduced without loss of 
exchange bias stability\cite{tsunodaJMMM06}.

Financial support was provided by the Hungarian National Scientific
Research Foundation (contract no. OTKA T068312, F68726 and NF061726) and by the EU via COST action P19, {\em Multiscale Modeling of Materials}.

\bibliographystyle{myst}


\begin{thebibliography}{10}

\bibitem{noguesJMMM99}
J. Nogu\'es and I.~K. Schuller, J. Magn. Magn. Mat. {\bf 192},  203  (1999).

\bibitem{nowakJMMM02}
U. Nowak, A. Misra, and K.~D. Usadel, J. Magn. Magn. Mat. {\bf 240},  243
  (2002).

\bibitem{miltenyiPRL00}
P. Milt\'enyi {\it et~al.}, Phys. Rev. Lett. {\bf 84},  4224  (2000).

\bibitem{vallejoAPL07}
G. Vallejo-Fernandez, L.~E. Fernandez-Outon, and K. O'Grady, Appl. Phys. Lett.
  {\bf 91},  212503  (2007).

\bibitem{aleyIEEE09}
N.~P. Aley {\it et~al.}, IEEE Trans. Magn.  submitted  .

\bibitem{mryasovEL05}
O.~N. Mryasov, U. Nowak, K. Guslienko, and R.~W. Chantrell, Europhys. Lett.
  {\bf 69},  805  (2005).

\bibitem{stauntonPRB06}
J. Staunton {\it et~al.}, Phys. Rev. B {\bf 74},  144411  (2006).

\bibitem{szunyoghPRB95}
L. Szunyogh, B. \'Ujfalussy, and P. Weinberger, Phys. Rev. B {\bf 51},  9552
  (1995).

\bibitem{voskoCJP80}
S.~H. Vosko, L. Wilk, and M. Nusair, Can. J. Phys. {\bf 58},  1200  (1980).

\bibitem{umetsuPRB04}
R.~Y. Umetsu, M. Miyakawa, K. Fukamichi, and A. Sakuma, Phys. Rev. B {\bf 69},
  104411  (2004).

\bibitem{tomenoJAP99}
I. Tomeno {\it et~al.}, J. Appl. Phys. {\bf 86},  3853  (1999).

\bibitem{sakumaPRB03}
A. Sakuma, K. Fukamichi, K. Sasao, and R.~Y. Umetsu, Phys. Rev. B {\bf 67},
  024420  (2003).

\bibitem{selteACS68}
K. Selte, A. Kjekshus, A.~F. Andresen, and W.~B. Pearson, Acta Chem. Scand.
  (1947-1973) {\bf 22},  3039  (1968).

\bibitem{udvardiPRB03}
L. Udvardi, L. Szunyogh, K. Palot\'as, and P. Weinberger, Phys. Rev. B {\bf
  68},  104436  (2003).

\bibitem{jansenPRB99}
H.~J.~F. Jansen, Phys. Rev. B {\bf 59},  4699  (1999).

\bibitem{umetsuAPL06}
R.~Y. Umetsu, A. Sakuma, and K. Fukamichi, Appl. Phys. Lett. {\bf 89},  052504
  (2006).

\bibitem{lichtensteinJMMM87}
A.~I. Lichtenstein, M.~I. Katsnelson, V.~P. Antropov, and V.~A. Gubanov, J.
  Magn. Magn. Mat. {\bf 67},  65  (1987).

\bibitem{nowakBOOK07}
U. Nowak,  in {\em Handbook of Magnetism and Advanced Magnetic Materials, Vol.
  2, Micromagnetism}, edited by H. Kronm\"uller and S. Parkin (John Wiley \&
  Sons Ltd., Chichester, 2007).

\bibitem{tsunodaJMMM06}
M. Tsunoda, K. Imakita, M. Naka, and M. Takahashi, J. Magn. Magn. Mat. {\bf
  304},  55  (2006).

\end{thebibliography}

\end{document}